\documentclass[prl,amsfonts,amssymb,twocolumn,showpacs]{revtex4-1}

\usepackage[breaklinks=true,colorlinks=true,linkcolor=blue,anchorcolor=blue,citecolor=blue,urlcolor=blue]{hyperref}
\usepackage{mathrsfs}
\usepackage{graphicx}

\def\hatd#1{\hat{#1}^\dagger}
\def\bra#1{\left\langle{#1}\right|}
\def\ket#1{\left|{#1}\right\rangle}
\def\slfrac#1#2{\left.#1\middle/#2\right.}

\begin{document}

%\begin{CJK*}{GBK}{song}

\title{Quantum Walks of Two Interacting Particles in One Dimension}

\author{Xizhou Qin$^{1}$}
\author{Yongguan Ke$^{1}$}
\author{Xiwen Guan$^{2,3}$}
\author{Zhibing Li$^{1}$}
\author{Natan Andrei$^{4}$}
\author{Chaohong Lee$^{1,}$}
\altaffiliation{Corresponding author.\\chleecn@gmail.com}

\affiliation{$^{1}$State Key Laboratory of Optoelectronic Materials and Technologies, School of Physics and Engineering, Sun Yat-Sen University, Guangzhou 510275, China}

\affiliation{$^{2}$State Key Laboratory of Magnetic Resonance and Atomic and Molecular Physics, Wuhan Institute of Physics and Mathematics, Chinese Academy of Sciences, Wuhan 430071, China}

\affiliation{$^{3}$Department of Theoretical Physics, Research School of Physics and Engineering, Australian National University, Canberra ACT 0200, Australia}

\affiliation{$^{4}$Department of Physics, Rutgers University, Piscataway, New Jersey 08854, USA}

\date{\today}

\begin{abstract}
We investigate continuous-time quantum walks of two indistinguishable particles (bosons, fermions or hard-core bosons) in one-dimensional lattices with nearest-neighbour interactions.
The two interacting particles can undergo independent- and/or co-walking dependent on both quantum statistics and interaction strength.
We find that two strongly interacting particles may form a bound state and then co-walk like a single composite particle with statistics-dependent propagation speed.
Such an effective single-particle picture of co-walking is analytically derived in the context of degenerate perturbation and the analytical results are well consistent with direct numerical simulation.
In addition to implementing universal quantum computation and observing bound states, two-particle quantum walks offer a novel route to detecting quantum statistics.
Our theoretical results can be examined in experiments of light propagations in two-dimensional waveguide arrays or spin-impurity dynamics of ultracold atoms in one-dimensional optical lattices.
\end{abstract}


\maketitle

Quantum walks (QWs)~\cite{Aharonov1993,Kempe2003}, the quantum counterpart of classical random walks (CRWs), have become a fundamental tool for developing quantum algorithms and implementing quantum computations.
In contrast to CRWs, which gradually approach to an equilibrium distribution, QWs spread ballistically if there is no decoherence.
The non-classical features of QWs offer versatile applications in quantum database search~\cite{Childs2004}, optimal element distinctness~\cite{Ambainis2007}, quantum simulation~\cite{Schreiber2012}, universal quantum computation~\cite{Childs2009,Childs2013}, and detection of topological states~\cite{Kitagawa2010,Kitagawa2010a,Kitagawa2012, Kraus2012, Verbin2013} and bound states~\cite{Kitagawa2012,Fukuhara2013a}.

Up to now, single-particle QWs have been implemented with several experimental systems.
In those experiments, the roles of quantum walkers are taken by single particles such as neutral atoms~\cite{Karski2009}, atomic ions~\cite{Schmitz2009,Zahringer2010}, photons~\cite{Schreiber2010, Broome2010, Schreiber2011}, and atomic spin impurities~\cite{Fukuhara2013}.
Attribute to their superpositions and interference features, single-particle QWs yield an exponential speedup over CRWs~\cite{Childs2003}.
However, it has been demonstrated that such an exponential speedup can be also achieved by classical waves~\cite{Knight2003,Perets2008}.

In contrast, by employing non-classical correlations, multi-particle QWs bring new benefits to practical quantum technologies.
For example, multi-particle QWs can implement universal quantum computations~\cite{Childs2013}.
In particular, two-particle QWs have been demonstrated via not only non-interacting photons in linear waveguide arrays~\cite{Hillery2010, Peruzzo2010, Lahini2010, Sansoni2012, Matthews2013, Meinecke2013} and but also interacting photons in nonlinear waveguide arrays~\cite{Solntsev2012,Lahini2012}.
Exotic quantum correlations have been observed even in the absence of inter-particle interactions~\cite{Bromberg2009, Peruzzo2010, Lahini2010, Matthews2013, Meinecke2013, Benedetti2012}.
Recently, the coexistence of free and bound states~\cite{Ganahl2012, Liu2013} has been dramatically observed through the QWs of two atomic spin-impurities in one-dimensional (1D) optical lattice~\cite{Fukuhara2013a}.
However, there is still lacking a comprehensive study of how QWs depend on quantum statistics and inter-particle interactions.
It is particularly interesting that how the co-walking of two interacting quantum walkers \emph{quantitatively} depend on their quantum statistics. Here, the co-walking of two walkers means their fully synchronized walking with same tendency and same speed.

In this Letter, we investigate throughout two-particle QWs in 1D lattices with nearest-neighbour interactions.
We explore how quantum statistics and inter-particle interactions competitively affect QWs.
In both position and momentum spaces, two-body correlations of bosonic and fermionic walkers show subtle bunching and anti-bunching signatures, respectively.
However, hard-core bosonic walkers show anti-bunching signature in position space and bunching signatures in momentum space.
For strong inter-particle interactions, we analytically derive the effective single-particle picture for the co-walking of two quantum walkers.
This presents a \emph{quantitatively} understanding for the effects of quantum statistics and inter-particle interactions in quantum co-walking.
The result of two hard-core bosonic walkers is consistent with the experimental observation of the two-magnon bound state~\cite{Fukuhara2013a}.
In the scenario of quantum-optical analogues~\cite{Longhi2009,Szameit2010}, the two-particle QWs can be experimentally simulated by light propagations in two-dimensional (2D) waveguide arrays~\cite{Szameit2009,Szameit2010,Corrielli2013}.

We consider QWs of two indistinguishable particles in 1D lattices described by the Hamiltonian with periodic boundary conditions,
\begin{equation}
  \hat H=-J\sum_{l=-L}^{L}{\left(\hat a_l^\dagger\hat a_{l+1}+\mathrm{h.c.}\right)}+V\sum_{l=-L}^{L}{\hat n_l\hat n_{l+1}}.\label{Hamiltonian}
\end{equation}
Here, $\hat a_l^\dagger$ ($\hat a_l$) creates (annihilates) a particle on the $l$-th lattice, $\hat n_l = \hat a_l^\dagger \hat a_{l}$ is the particle number, $J$ is the nearest-neighbour hopping, and $V$ stands for the nearest-neighbour interaction.
The Hamiltonian~(\ref{Hamiltonian}) associates with the quasi-particle representation for a XXZ Heisenberg chain~\cite{Matsubara1956,Dziarmaga2005}.
The propagation dynamics in two-particle systems ($\hat{N}=\sum_{l=-L}^{L}{\hat n_l}=2$) represents a class of continuous time two-particle QWs.

We consider three different types of commutation relations (CRs) for the particle operators: bosonic, fermionic and hard-core bosonic ones.
The bosonic CRs read $[\hat a_l,\hat a_k]=[\hat a_l^\dagger,\hat a_k^\dagger]=0$ and $[\hat a_l,\hat a_k^\dagger]=\delta_{lk}$.
The fermionic CRs present $\{\hat a_l,\hat a_k\}=\{\hat a_l^\dagger,\hat a_k^\dagger\}=0$ and $\{\hat a_l,\hat a_k^\dagger\}=\delta_{lk}$.
The hard-core bosonic CRs give $[\hat a_l,\hat a_k]=[\hat a_l^\dagger,\hat a_k^\dagger]=[\hat a_l,\hat a_k^\dagger]=0$ for $l\neq{k}$, while $\{\hat a_l,\hat a_l\}=\{\hat a_l^\dagger,\hat a_l^\dagger\}=0$ and $\{\hat a_l,\hat a_l^\dagger\}=1$.
By implementing a discrete Fourier transformation~\cite{Javanainen2010} $\hat a_\alpha^\dagger={1 \over \sqrt{L_t}} \sum_{l=-L}^{L}e^{-ip_\alpha l}\hat a_l^\dagger$ [where the quasi-momentum $p_\alpha=2\pi\alpha/L_t$, the integer $\alpha=\left(-L,-L+1,\cdots,0,\cdots,L-1,L\right)$ and the total lattice number $L_t=2L+1$], one can obtain CRs in momentum space: (1) $[\hat a_\alpha,\hat a_\beta]=[\hat a_\alpha^\dagger,\hat a_\beta^\dagger]=0$ and $[\hat a_\alpha,\hat a_\beta^\dagger]=\delta_{\alpha\beta}$ for bosons; (2) $\{\hat a_\alpha,\hat a_\beta\}=\{\hat a_\alpha^\dagger,\hat a_\beta^\dagger\}=0$ and $\{\hat a_\alpha,\hat a_\beta^\dagger\}=\delta_{\alpha\beta}$ for fermions; and (3) $[\hat a_\alpha,\hat a_\beta]=[\hat a_\alpha^\dagger,\hat a_\beta^\dagger]=0$ and $[\hat a_\alpha,\hat a_\beta^\dagger]=\delta_{\alpha\beta}-{2 \over L_t}\sum_{\gamma\gamma'}\delta_{[\alpha+\gamma,\beta+\gamma']}\hat a_\gamma^\dagger\hat a_{\gamma'}$ for hard-core bosons (HCBs).
The system of HCBs is equivalent to a XXZ Heisenberg chain~\cite{Matsubara1956}, which can be realized by ultracold two-level atoms in optical lattices~\cite{Duan2003, Kuklov2003, Garcia-Ripoll2003, Altman2003, Lee2004, Fukuhara2013, Fukuhara2013a}.

We now discuss the Hilbert space involved by the two-particle QWs.
Since $[\hat{N}, \hat{H}]=0$, the total particle number $N$ is conserved and the system will evolve in the two-particle Hilbert space.
For two bosons, their Hilbert space is spanned by basis, $\mathcal{B}^{(2)}_{\mathrm{B}}=\left\{\ket{l_1 l_2} = (1+\delta_{l_1 l_2})^{-\frac{1}{2}} \hatd a_{l_1}\hatd a_{l_2}\ket{\mathbf{0}}, -L\leq l_1 \leq l_2\leq L\right\}$.
For two fermions or two HCBs, their Hilbert spaces are spanned by the same basis, $\mathcal{B}^{(2)}_{\mathrm{FH}}=\left\{\ket{l_1 l_2}=\hatd a_{l_1}\hatd a_{l_2}\ket{\mathbf{0}},-L \leq l_1<l_2 \leq L\right\}$. Given $\mathcal{B}^{(2)}_{\mathrm{B}}$ and $\mathcal{B}^{(2)}_{\mathrm{FH}}$, it is easy to find the Hamiltonian matrix $H^{(2)}$ in two-particle sector.
In units of $\hbar=1$, the time evolution of an arbitrary state obeys
\begin{equation}
i\frac{d}{dt}\left|\psi(t)\right\rangle = H^{(2)} \left|\psi(t)\right\rangle,\label{TE}
\end{equation}
with $\left|\psi(t)\right\rangle=\sum_{l_1 \le l_2} C_{l_1, l_2} (t) \ket{l_1 l_2}$ for bosons and $\left|\psi(t)\right\rangle=\sum_{l_1 < l_2} C_{l_1, l_2} (t) \ket{l_1 l_2}$ for fermions and HCBs.

Below we will study two-particle QWs starting from an initial state $\left|\psi_\mathrm{ini}\right\rangle=\hat a_0^\dagger\hat a_1^\dagger\left|\mathbf{0}\right\rangle$.
Here, $\left|\mathbf{0}\right\rangle$ denotes the vacuum state.
In order to explore the correlation between two quantum walkers, we calculate the two-particle correlation in position space,
\begin{equation}
  \Gamma_{qr}=\bra{\psi(t)}\hatd a_q\hatd a_r\hat a_r\hat a_q\ket{\psi(t)},
\end{equation}
and the ones in momentum space,
\begin{equation}
  \Gamma_{\alpha\beta} = \bra{\psi(t)}\hatd a_\alpha\hatd a_\beta\hat a_\beta\hat a_\alpha\ket{\psi(t)},
\end{equation}
with $\left|\psi (t) \right\rangle$ given by Eq.~(\ref{TE}).
The two-particle correlation in position and momentum spaces for difference quantum statistics and interaction strength provide a clear insight into the two-particle QWs, see Figs.~1 and 2.

\begin{figure}[!htp]
\includegraphics[width=1.0\columnwidth]{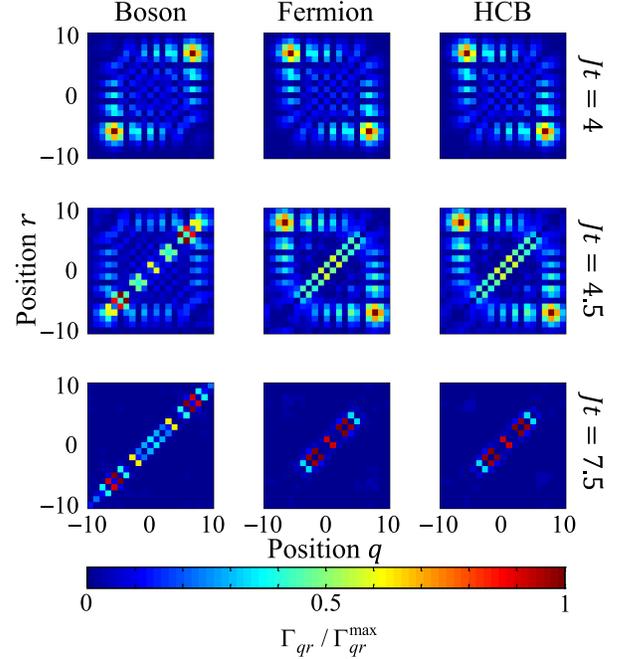}
\caption{\label{Fig_TPCF_Position}(Color online) Two-particle correlations of quantum walkers in position space. The first, second and third columns correspond to Bose, Fermi and HCB statistics, respectively. The interaction strength $\left|V/(2J)\right|=0,\,0.5$, and $2$ for the first, second and third rows, respectively. Here we only show the instantaneous correlations before colliding with the boundaries $l=\pm 10$, whose evolution times are given by $Jt=4$, $4.5$ and $7.5$, respectively.}
\end{figure}

In the position space, the correlations of two bosonic walkers (the first column of Fig.~1) show the bunching behavior, while the correlations of two fermionic walkers (the second column of Fig.~1) and two hard-core bosonic walkers (the third column of Fig.~1) show anti-bunching behavior.
We observe that the correlations of fermions and HCBs in the position space almost have no difference.
This is because that, a spin-$\frac{1}{2}$ Heienberg XXZ model, which is equivalent to a hard-core Bose-Hubbard model~\cite{Matsubara1956}, can be mapped onto a Hubbard-like model of spinless fermions via Jordan-Wigner transformation~\cite{Dziarmaga2005}.
Although boundary conditions of the Hubbard-like model of spinless fermions depend on the total particle number~\cite{Dziarmaga2005}, the boundary conditions have no effect on the dynamics before the two walkers hit the boundaries.
Therefore the correlations are almost the same for fermions and HCBs.
Consequently, from bunching and anti-bunching of the two quantum walkers in position space, one can distinguish Bose statistics from Fermi and HCB ones.

\begin{figure}[!htp]
\includegraphics[width=1.0\columnwidth]{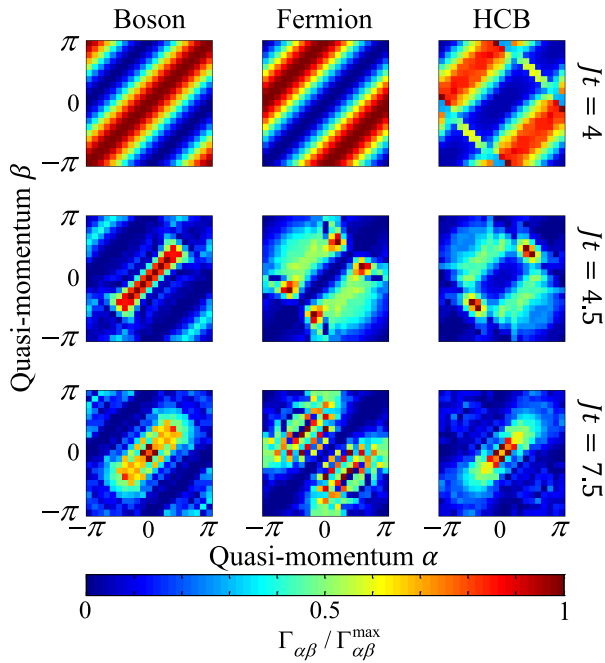}
\caption{\label{Fig_TPCF_Momentum}(Color online) Two-particle correlations in momentum space correspond to the ones shown in Fig.~1.}
\end{figure}

On the other hand, the correlations of bosonic and hard-core bosonic walkers in momentum space show bunching behavior, see the first and third columns of Fig.~2.
Nevertheless the correlations of fermionic walkers (the second column of Fig.~2) show anti-bunching behavior.
This means that bunching and anti-bunching of the two quantum walkers in momentum space reveal different features of fermions and HCBs.
Therefore, from bunching and anti-bunching of the two quantum walkers in both position and momentum spaces, one can distinguish the three statistics: Bose, Fermi and HCB ones.

The correlations $\Gamma_{qr}$ on the minor diagonal lines $(q=r \pm 1)$ are enhanced when the interaction-hopping ratio increases, see Fig.~1.
Since $\Gamma_{q,q\pm 1}$ presents a joint probability of finding one walker on the $q$-th site and the other walker on the $(q\pm 1)$-th site, the significant correlations on the minor diagonal lines is a robust signature of co-walking.
The co-walking is also an important signature of the existence of two-particle bound states, see~\cite{Fukuhara2013a, Ganahl2012, Liu2013} for the case of two magnons.
Usually, two interacting quantum walkers simultaneously undergo independent- and co-walking.

\begin{figure}[!htp]
\includegraphics[width=1.0\columnwidth]{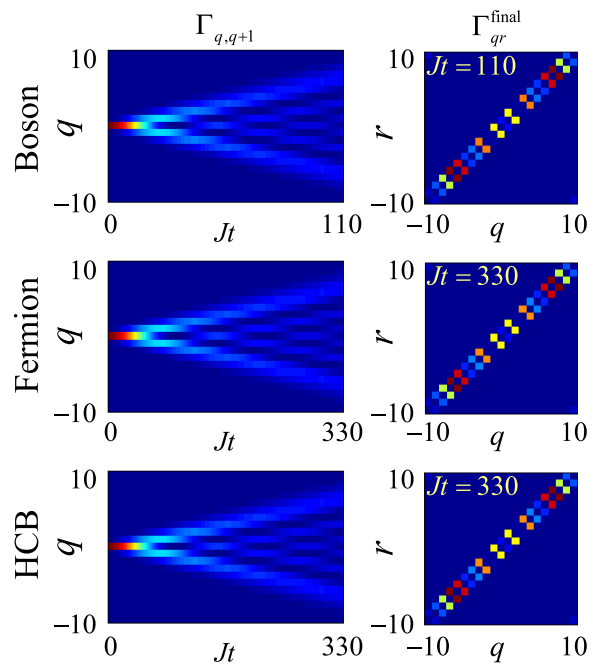}
\caption{\label{Fig_TPCF_Bound}(Color online) Co-walking of two strongly interacting walkers with $\left|V/(2J)\right|=40$. Left: Time evolution of the minor diagonal correlations $\Gamma_{q,q+1}$. Right: Two-particle correlations $\Gamma_{q,r}^{final}$ for the final states.}
\end{figure}

For strong inter-particle interaction, i.e. $\left|V/J\right| \gg 1$, the two quantum walkers behave like a single composite particle and their QWs are dominated by co-walking.
We thus can treat the hopping term as a perturbation to the interaction term in the Hamiltonian~(\ref{Hamiltonian}).
By employing the second-order perturbation theory for degenerate systems~\cite{Takahashi1977}, we analytical obtain an effective single-particle model for the co-walking of the two quantum walkers.
If the two walkers start from two neighbour lattice sites, their co-walking dynamics can be described by superposition of states $\ket{n_q=1,n_{q+1}=1}=\hatd a_q\hatd a_{q+1}\ket{\mathbf{0}}$ with different $q$ (where $q=-L,-L+1,\cdots,0,\cdots,L-1,L$).
During the process of co-walking, the two particles behave like a single composite particle.
In order to capture the single-particle nature of the co-walking, we introduce a creation operators $\hatd b_q$ for the composite particle consisting of one particle on the $q\textrm{-th}$ lattice site and the other particle on the $\left(q+1\right)\textrm{-th}$ lattice site.
Explicitly, $\hatd b_q \Leftrightarrow \hatd a_q\hatd a_{q+1}$ and $\ket{n^c_q=1}=\hatd b_q\ket{\mathbf{0}} \Leftrightarrow \ket{n_q=1,n_{q+1}=1}=\hatd a_q\hatd a_{q+1}\ket{\mathbf{0}}$.
Therefore, the two bosonic walkers obey an effective single-particle Hamiltonian,
\begin{equation}
  \hat H^{(2)}_\mathrm{eff} = J^\mathrm{B}_\mathrm{eff} \sum_q\left(\hatd b_q\hat b_{q+1} +\hatd b_{q+1}\hat b_q\right)+\mu^\mathrm{B}_\mathrm{eff} \sum_q\hatd b_q\hat b_q,
\end{equation}
with the hopping strength $J^\mathrm{B}_\mathrm{eff}=\slfrac{3J^2}{V}$ and the chemical potential $\mu^\mathrm{B}_\mathrm{eff}=V+\slfrac{6J^2}{V}$.
Similarly, the two fermionic walkers and the two hard-core bosonic walkers obey the same effective single-particle Hamiltonian
\begin{equation}
  \hat H^{(2)}_\mathrm{eff} = J^\mathrm{FH}_\mathrm{eff} \sum_q\left(\hatd b_q\hat b_{q+1} +\hatd b_{q+1}\hat b_q\right) +\mu^\mathrm{FH}_\mathrm{eff} \sum_q\hatd b_q\hat b_q,
\end{equation}
with $J^\mathrm{FH}_\mathrm{eff}=\slfrac{J^2}{V}$ and $\mu^\mathrm{FH}_\mathrm{eff}=V+\slfrac{2J^2}{V}$.

We observe that, for fixed values of $J$ and $V$, the hopping strength ($J^\mathrm{B}_\mathrm{eff},\, J^\mathrm{FH}_\mathrm{eff}$) of the composite particle essentially depend on their quantum statistics.
This means that quantum statistics has a significant effect on the co-walking of two interacting walkers.
Fig.~\ref{Fig_TPCF_Bound} shows numerical simulations with the minor diagonal correlations $\Gamma_{q,q+1}$ and the two-particle correlations $\Gamma_{q,r}^{final}$ for the final states with $\left|V/(2J)\right|=40$.
From the correlations $\Gamma_{q,r}^{final}$ in the right column of Fig.~\ref{Fig_TPCF_Bound}, we find that the two strongly interacting walkers are dominated by co-walking.
From the time evolution of $\Gamma_{q,q+1}$ in the left column of Fig.~\ref{Fig_TPCF_Bound}, we see that the propagation speed of two bosonic walkers is three times of the ones of the fermionic and hard-core bosonic walkers.
These numerical results of propagation speeds are well consistent with our analytical prediction $J^\mathrm{B}_\mathrm{eff} = 3J^\mathrm{FH}_\mathrm{eff}$ from the second-order perturbation theory.

\begin{figure}[!htp]
\includegraphics[width=1.0\columnwidth]{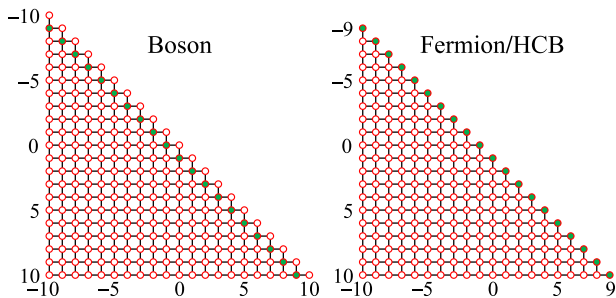}
\caption{\label{Fig_Waveguide}(Color online) Classical simulation with two-dimensional optical waveguide arrays. Each circle represents a waveguide. Green-colored and uncolored circles label waveguides with different refractive indices. The black lines connecting different circles denote their couplings. Left: the waveguide arrays for simulating two bosons. Right: the waveguide arrays for simulating two fermions or two hard-core bosons.}
\end{figure}

The two interacting quantum walkers in our models can be experimentally simulated with ultracold atoms in optical lattices and light waves in waveguides.
By using spin impurities of ultracold atoms in optical lattices,
two-magnon dynamics in a 1D Heisenberg XXZ chain has been observed in a recent experiment~\cite{Fukuhara2013a}.
It was a dramatic realization of two-HCB quantum walks with intermediate interaction ($\Delta=\left|V/(2J)\right|=0.986$).
The strong interaction regime ($\Delta \gg 1$) can be achieved by Feshbach resonance~\cite{Widera2004,Gross2010}.
Moreover, based on the quantum-optical analogues using engineered photonic waveguides~\cite{Longhi2009,Szameit2010}, the two-particle QWs obeying the Hamiltonian~(\ref{Hamiltonian}) can be simulated via light propagations.
As a single quantum walker in a 2D lattice is equivalent to two quantum walkers in a 1D lattice~\cite{Schreiber2012}, the two-particle QWs in 1D lattices can be simulated with light waves in 2D waveguide arrays~\cite{Szameit2009,Szameit2010,Corrielli2013}.
The temporal evolution of the superposition amplitude $C_{l_1,l_2}$ in the two-particle Hilbert space is mapped onto the spatial propagation of the optical field $E_{l_1,l_2}$ in the $(l_1,l_2)$-th waveguide.
According to the evolution equation~(\ref{TE}) of $C_{l_1,l_2}$, the propagation equation for $E_{l_1,l_2}$ is given by
\begin{eqnarray}
i\frac{d}{d z} E_{l_1,l_2} &=& - J\left(E_{l_1,l_2+1} +E_{l_1,l_2-1}\right) \\
&&-J\left(E_{l_1+1,l_2} +E_{l_1-1,l_2}\right)+V_{l_1, l_2} E_{l_1,l_2},\;\nonumber
\end{eqnarray}
with $V_{l_1, l_2}=V\delta_{l_1, l_2\pm 1}$ and the propagation distance $z$.
In Fig.~\ref{Fig_Waveguide}, we shown the 2D waveguide arrays for simulating two-particle QWs with $L=10$.
Similar to the 2D waveguide arrays used in recent experiments~\cite{Szameit2009,Szameit2010,Corrielli2013}, the waveguide arrays shown in Fig.~\ref{Fig_Waveguide} can be fabricated in a silica substrate by direct waveguide writing with femtosecond lasers~\cite{Gattass2008}.
Here the inter-particle interaction strength $V$ is controlled by the difference of refractive indices between green-colored and uncolored waveguides, see Fig.~\ref{Fig_Waveguide}.

In summary, we have studied how quantum statistics and inter-particle interaction affect two-particle QWs in 1D lattices with nearest-neighbour interactions.
Due to the inter-particle interaction, two particles with different quantum statistics undergo independent- and/or co-walking.
The QWS are dominated by independent-walking in the weak interaction limit, and vice versa, they are dominated by co-walking in the strong interaction limit.
We have analytically derived the effective single-particle model for the co-walking of two strongly interacting particles.
We find that the propagation speed for the co-walking of two bosons is exactly three times of the ones for the co-walking of two fermions or two HCBs.
Our results for the case of two HCBs are well consistent with the recent experimental observation of quantum dynamics of two atomic spin impurities~\cite{Fukuhara2013a}.
We have also demonstrated that the two-particle QWs in our models can be simulated by light propagations in engineered 2D waveguide arrays~\cite{Szameit2009,Szameit2010,Corrielli2013}.
Besides implementing universal quantum computation~\cite{Childs2009,Childs2013} and observing bound states~\cite{Kitagawa2012,Fukuhara2013a}, our results of two-particle QWs provide promising applications in detecting quantum statistics.

We thank Gora Shlyapnikov for discussion. This work is supported by the NBRPC under Grants No. 2012CB821305 and 2012CB922101, the NNSFC under Grants No. 11374375 and 11374331, the Ph.D. Programs Foundation of Ministry of Education of China under Grant No. 20120171110022, and the NCETPC under Grant No. NCET-10-0850. XWG is partially supported by the Australian Research Council.

%

%\end{CJK*}

\end{document}